# Self-assisted Amoeboid Navigation in Complex Environments

Inbal Hecht[(1),#], Herbert Levine[(2)], Wouter-Jan Rappel[(2)] and Eshel Ben-Jacob[(1),(2),#]

[(1)]The Sackler School of Physics and Astronomy, Tel Aviv University, P.O. Box 39040, Tel Aviv 69978, Israel
[(2)]The Center for Theoretical Biological Physics, University of California San Diego, La Jolla, California 92093

[#]**Corresponding authors: inbal.hecht@gmail.com ; eshelbj@gmail.com**

## Abstract

**Background:** Living cells of many types need to move in response to external stimuli in order to accomplish their functional tasks; these tasks range from wound healing to immune response to fertilization. While the directional motion is typically dictated by an external signal, the actual motility is also restricted by physical constraints, such as the presence of other cells and the extracellular matrix. The ability to successfully navigate in the presence of obstacles is not only essential for organisms, but might prove relevant in the study of autonomous robotic motion.

**Methodology/principal findings:** We study a computational model of amoeboid chemotactic navigation under differing conditions, from motion in an obstacle-free environment to navigation between obstacles and finally to moving in a maze. We use the maze as a simple stand-in for a motion task with severe constraints, as might be expected in dense extracellular matrix. Whereas agents using simple chemotaxis can successfully navigate around small obstacles, the presence of large barriers can often lead to agent trapping. We further show that employing a simple memory mechanism, namely secretion of a repulsive chemical by the agent, helps the agent escape from such trapping.

**Conclusions/significance:** Our main conclusion is that cells employing simple chemotactic strategies will often be unable to navigate through maze-like geometries, but a simple chemical marker mechanism (which we refer to as "self-assistance") significantly improves success rates. This realization provides important insights into mechanisms that might be employed by real cells migrating in complex environments as well as clues for the design of robotic navigation strategies. The results can be extended to more complicated multi-cellular systems and can be used in the study of mammalian cell migration and cancer metastasis.

## Introduction

Cellular migration has been an intriguing phenomenon for many years. From wound healing and immune response of mammalian cells, to chemotaxing bacteria and amoeba, living cells exhibit a variety of motility abilities [1-3].

Most motile cells attempt to follow external directional signals (in the form of chemical or mechanical gradients) while moving in a complex environment. For example, immune system cells follow chemical gradients as they leave the vascular system and migrate through cellular tissues towards the site of an inflammation [1-3], and metastasizing cancer cells invade through the surrounding tissues to form secondary tumors [4,5]. These processes require transiting through an environment consisting of other cells and extracellular matrix (ECM). The chemotactic process therefore involves both a response to the external signal, and the handling of mechanical constraints on the motion.

From the computational point of view, much research has been devoted to the study of autonomous motion planning [6]. An important part of autonomous taxis is the ability to independently navigate, namely to find a path to a defined target under possible constraints. This ability, which is essential for cellular translocation [7-10] and for the study of animal behavior [11], is also important for successful robotic exploration. For individual agent-based navigation, one obvious way of encoding target information is by having the target emit a signal (steady or dynamic), which allows the agent to determine a locally favorable direction. But, it is clear that the locally best direction may not be the overall best choice, as this may lead to trapping of the agent by large obstacles. Optimally, the agent should balance this target-based information with local structural information so as to navigate around these traps. The conceptual view that cells should integrate multiple sources of information can lead to new predictions regarding cellular chemotaxis; this will be seen below.

In this work, we will study these questions by use of a simplified model of cellular navigation capabilities. Efforts in the biological and biophysical community have elucidated the basic elements underlying how cells are able to navigate via chemical gradients. First, the external signal influences the cell orientation by various signal transduction pathways, highly conserved between different cell types. Consequently, the cell polarizes and different chemicals accumulate at the front versus the back of the cell. Motility is typically obtained by f-actin polymerization at the cell's front, leading to membrane protrusions such as pseudopods, lamellipods and ruffles. Beyond individual propulsion, the cell interacts with its environment by various passive as well as active processes: The cell can adhere or de-adhere to the extra -cellular matrix or to neighboring cells [12,13], apply forces and even actively degrade the ECM by proteases [14].

Many attempts have been made to model different aspects of directed migration and chemotaxis. Most models to date have addressed distinct parts of motility, including retraction and protrusion [15], but are unable to describe the entire motility process; other models use ad-hoc rules to describe the motion [16]. Many studies, both theoretical and experimental, have also been devoted to the question of collective motion and how it emerges from individual interactions [17-19], from the cellular to the animal scale. In this work we focus on single cell motility, but our results can be extended to the case of collective motion by adding intra-cellular interactions.

Here, as a first step in the study of navigation in complex media, we study the ability of a moving cell to navigate between obstacles. We isolate the sensing and motility from adhesion and proteolysis and focus on the strategies needed by the cell so as to find its way under environmental constraints. We do so by using a simulated amoeboid, crawling in different environments according to an external signal. We first study the characteristics of free motion in a chemoattractant gradient, then turn to the effects of obstacles in the medium, and finally to the more challenging case of navigation in a maze. In the case of a maze, motion according to the local chemical direction can cause the cell to become trapped by the maze walls. When this occurs, the cell needs to retrace its steps, moving away from the optimal chemical direction, in order to find a new pathway and resume its motion towards the target.

We demonstrate that memory-less navigation yields very low success rates, and that in most cases the cell becomes stuck in a maze corner or dead end, and cannot reach its goal. We then show that a simple memory effect mediated by a chemical marker secreted and detected by the cell, can lead to much higher success rates. We propose to term this type of behavior "self-assisted" navigation since the cell by virtue of the marker emission is able to recognize that it is trapped and thereby alter its behavior so as to assist itself in trying to escape. We hypothesize that navigation based on this type of mechanism is a likely possibility for chemotaxis in complex environments; this can obviously be tested in, for example, microfluidics devices where flow can be used to interfere with marking strategies. Finally, this finding provides insights into needed components for successful robotic motion planning.

## Results

### Amoeboid motion

Amoeboids, unlike bacteria, can directly detect spatial gradients in chemical concentration, responding to as low as a 2% difference in concentration between the cell's front and back [20-22]. Chemotaxis, i.e. motion according to the gradient direction, is then achieved by sending out membrane protrusions (pseudopods), with a typical life time. Pseudopods are mostly created in the leading edge of the cell (cell's front), but some pseudopods may emerge also from the sides, depending on the gradient strength and cell polarization [23,24]. The overall cellular motion is achieved by retraction of the cell's rear towards the advancing front. Pseudopods typically exhibit complex behavior of bifurcation and retraction, with some periodicity of right-left split directions [25].

The formation of pseudopods is accompanied by accumulation of various effectors on the membrane, in the form of "patches" with limited lifetime [26]. These patches were shown to spatially correlate with the location of pseudopods [27,28]. Genetic studies have verified that these effectors are controlled by the external chemical signal and in turn are responsible for activating the machinery that drives the extensional dynamics.

**Amoeboid directional motion: Model**

In our model, the internal direction of the cell (the cell polarization axis) is determined by the external gradient of the signaling field with some added noise. This simple mechanism is intended to mimic the gradient sensing process, without explicitly dealing with receptor occupancies. And indeed, in our previous work [29] we have shown that such a "noisy compass" mechanism can produce pseudopod statistics and overall directional motion which closely resemble real cell's behavior.

The compass noise is drawn from a Gaussian distribution, and the distribution width is inversely related to the steepness of the gradient: A steep gradient yields more accurate directional sensing, and hence there is less noise (and vice versa); this was chosen so that our model would be in agreement with experimental data comparing response at different gradient strengths [29]. The membrane point which is the closest to the internal gradient vector is chosen to be the new cell front. This directional sensing process takes place every few minutes and has no hysteresis.

After determining the new front position, a patch of activation is created on the membrane. This patch determines the membrane area that will be pushed outward to create a pseudopod. The size and lifetime of the patch are randomly drawn from a given range, fitted to experimental data (See Supporting Table S1 for more details). After this lifetime, the patch is gradually degraded and a new patch is created. Other forces acting on the cell membrane include cortical tension, constant area constraint and friction (see Supporting Text S1 for more details). The forward front pushing and back retraction, due to the constant area constraint, lead to net forward motion. The cortical tension determines the width of membrane protrusions and influences the shape of the cell (e.g. round or slender). Once all the forces are calculated, we determine the velocity change at each membrane point and advance the membrane simultaneously. Technical and computational details, and the form of each of the forces, are given in detail in Supporting Text S1.

**Freely moving amoeboid**

When no obstacles are present, the cell motion and shape dynamics depend on the internal parameters such as the activation patch size, patch lifetime and the ratio between the protruding force and the cortical tension, as well as on the steepness of the gradient. Generally, a cell can have several, independent areas of activation ('patches'), and the number of protrusions will vary accordingly. In Fig. 1 we show the simulated movement of two typical model cells, one with a single activated patch (Fig 1.(a)-(b)) and one with two activated patches (Fig. 1(c)-(d)). Different types of cells exhibit different numbers of protrusions, and the number of patches in the model can be tuned to match a specific cell type. In this work we focus on the case of a single protrusion, mostly for simplicity and to save computation time.

The chemotactic index (CI) of the cell is defined as the ratio between the distance traveled in the direction of the signal and the overall distance traveled by the cell, and experimentally was found to depend on the gradient steepness [24]. In our model, steepness lowers the noise variance and for a narrower noise distribution, the CI increases and the cell path becomes more accurate. In Fig. 2 we show typical paths for the cases of low versus high noise levels. It is easy to see that the path to the target is more accurate when the noise level is low, namely when the gradient is steeper. This has been studied in detail elsewhere, using a more complex reaction-diffusion model for generating the patches [24].

**Navigation between obstacles**

Amoeboid-like motion is directly advantageous in the presence of obstacles. When the cell encounters an obstacle directly ahead, it cannot continue to move in the direction of the protrusion, but can still move in other directions. As a result, the cell slides along the walls of the obstacle, as the points adjacent to the obstacle are stuck while points slightly away are free to move. When a new activation patch is created, it will typically not be centered at a point that is attached to the obstacle; rather it will be shifted to a free point. This simple mechanism yields an efficient obstacle-passing mechanism, as can be seen in Fig. 3. It also creates an impression of the obstacles "guiding" the motion, as seen in experiments [29,30]. It is the nature of amoeboid motion, i.e. the creation of stochastic and recurring protrusions, as well as a flexible cell shape, that allows for this efficient obstacle circumvention. Ongoing directional change is a constant feature of amoeboid motion, and thus the process of bypassing an obstacle does not demand any additional special mechanism or procedure. It should be noted, though, that this notion only applies to obstacles of the amoeboid's size or less, as will be shown in the next section. Navigation between obstacles is mostly a question of locally bypassing one obstacle at a time. As long as the cell is flexible enough and the obstacles are not too large, the cell can fit between the obstacles and proceed forward without the need for sophisticated longer-range analysis, memory or "intelligence".

**Navigation in a maze**

The situation changes when the cell is exposed to a complex terrain with obstacles larger than the cell size. In this case the cell may spend a long time in attempts to bypass the obstacles, especially when they happen to be perpendicular to the direction of the chemical gradient. We have chosen to illustrate this situation by considering navigation in a maze, as the cell can now be trapped by the maze walls. The example shown in Fig. 4 presents a case in which the chemical gradient points to the central top area of the maze, with several possible pathways from the starting point to this target. Importantly, signaling molecules can freely diffuse through the maze walls (see also the Discussion section below), and as a result, there are points inside the maze where the local direction of the gradient leads to a corner or a dead end (marked with asterisks in Fig. 4). A chemotaxing amoeboid that precisely follows the signal may therefore get stuck in such a local trap. Escaping the trap demands motion in a direction different from the one dictated by the signal. This poses a challenge that may demand more than the simple chemotaxis capability that is sufficient in the case of small obstacle circumvention.

When the noise distribution width is taken to be large (corresponding in our model to the case of a small gradient) and fixed, the cell's path is curved and the cell can indeed explore different paths in the maze; this is shown in Fig. 5(a)-(b). With high noise the cell may get off the trail, but it can also overcome a local barrier by moving in an opposite direction for a short while. This behavior is not all that beneficial to the cell though, since the search is inefficient; the cell spends a long time en route and sometimes wanders far away from the well-defined external direction. When the noise is lower, the cell always chooses the shortest path and wanders around much less, as shown in Fig. 5(c), but is also easily stuck (**Error! Reference source not found.**(d)) as it obeys the external constraint precisely and cannot move against the dictated direction

**Adaptive noise**

One possible strategy to evade traps is that of noise adaptation. In our baseline model, the noise has a fixed value, set by the global strength of the gradient. However, organisms moving in different environments may need to adapt their noise level and adjust it to the different terrains. In fact, chemotaxing amoebas such as *Dictyostelium* automatically exhibit different noise levels when moving in different strength local gradients [31], due to the underlying mechanism of directional sensing by receptor occupancy differences. Following this notion, we choose the noise distribution width to *dynamically* depend on the relative difference in the chemoattractant concentration between the cell's front and back:

$$\sigma = \frac{C_{min}}{C_{max} - C_{min}},$$

where the minimal concentration defines the cell back and the maximal concentration defines the cell front. A large difference between the cell's front and back results in a narrow noise distribution, while a small difference between the front and the back implies noisier directional sensing and a wider noise distribution. This dynamic noise selection allows the cell to adjust to the varying environment and partially optimize its search strategy.

However, even with this type of adaptive noise, chemotactic navigation is not very successful. To estimate the efficiency and success rate of each of the tested strategies, we repeated the simulation in the same maze and with the same starting point. Given the stochasticity of the patch generation process in the model, a different path was obtained in each run. If the cell managed to get to the defined ending point within the defined simulation time limit, it is counted as successful. Out of approximately 250 such runs, the computational cell was found to become stuck in 70.1% of the cases and only 29.9% of the cells could eventually reach the target (red zone). In two other maze/signal configurations, 100% of the cells became stuck. An example of a trapped cell and its path is shown in Fig. 6(a). Therefore, navigating via the chemoattractant gradient by itself is insufficient in the case of large obstacles that block the way. Motion against the gradient on a scale larger than the cell's length is needed.

**The "self-assistance" strategy**

To consider a possible new navigation strategy that can enhance escape from traps, we add to our model a repulsive chemical field, continuously secreted by the cell. When the cell is trapped in a specific location, the level of this chemical increases and acts to mask the external chemoattractant. For simplicity, we assumed that the two chemicals (i.e. the external chemoattractant and the secreted chemorepellent) have exactly opposite effects on the cell's navigation, so the cell observes an effective field given by:

$$C_{eff} = C_{sig} - C_{chem},$$

where $C_{sig}$ is the concentration of the external signal, $C_{chem}$ is the concentration of the secreted chemical, and $C_{eff}$ is the effective concentration detected by the cell. The secretion and detection of this chemical marker acts as a memory, which actively marks areas in the maze that have already been visited. This chemical can diffuse, with various possible choices for the diffusion rate; as we will see, the only significant limitation is that it cannot diffuse too quickly and thereby lose the information as to the cell's recent positions. We term this strategy "self-assistance", as the cell is able to escape the trap without any external help. The effect of this augmented navigation is demonstrated in Fig. 6(b) (the full list of parameters values is given in Supporting Table S1). The cell is initially trapped by the maze walls (as in Fig. 6(a)) but due to the increasing level of secreted repulsive chemical, the cell eventually selects a new direction that leads it out of the corner. Additional details regarding this effect is shown in Fig. 6(c), where the difference between the chemoattractant and the chemorepellent is represented using a color code. The areas that had been previously visited by the cell have a clearly lower effective concentration, and the color roughly indicates the time spent in a specific location. The success rate for this type of navigation is significantly higher than that of simple chemotaxis – using around 250 runs on the maze shown in Fig. 4, the success rate rose from 29.9% to as high as 99%. For the other maze/signal configurations that were tested we obtained an increase from zero success to 72% and 76% using the "self-assistance" procedure (data not shown).

We specifically tested the effect of the repulsive chemical's diffusion on the success rates of the moving cell. As expected, the success rate is high for low diffusivities, and falls significantly when diffusion rate exceeds a threshold, as shown in Fig. **7**. This is reasonable, as fast diffusion blurs the spatial information needed to make the correct decision and instead the cell actually responds only to the external chemical gradient .

The passage time in the maze, namely the time it takes to reach the target, is an indicator of search efficiency. The search time for "self-asssisted" navigation is significantly shorter than that of a simple chemotactic navigation (see Fig. 8). Just as we found for the success rate, the search time increases rapidly (namely the efficiency is decreased) as the chemical diffusion rate exceeds the threshold.

## Discussion

Amoeboid motion is found in different biological systems, not only in amoeba per se but also in migrating mammalian cells. For example, metastasizing cancer cells are able to transform from the mesenchymal mode of motion, in which they move by degradation of the ECM, to the amoeboid mode, in which they move through the ECM by cell shape deformations [24]. Invading cells, both normal and cancerous, interact with the extracellular matrix (ECM) in a two-fold manner. The ECM acts as a physical scaffold that binds cells together into tissues and guides cellular migration along matrix tracks, but it also presents a physical barrier that cells need to overcome – either by proteolysis or by a shape change. The roles of the ECM and the influence of its specific features (such as pore size and density) on cellular invasion are in the focus of current biomedical research, both

*in vivo* and *in vitro* [6,32]. Importantly, the ECM also elicits biochemical and biophysical signaling, which may influence cellular differentiation and motility.

Migrating amoeboid cells are able to change their shape drastically in order to adapt to encountered constraints and thereby push through narrow places. These cells have low levels of integrin expression, reduced focal contacts, a low degree of adhesiveness and significantly higher motility velocities as compared to mesenchymal cells [33,34]. Amoeboid shape-driven migration allows cells to evade, rather than degrade, barriers, and enables migration even when mesenchymal motility is impossible. Recent experimental work has shown that cells that were restricted to amoeboid motility, by inhibition of matrix metalloproteases (MMPs) could still invade pores that were as small as the nuclear size of the cell [35-40]. New evidence also supports a central role for amoeboid motility in cell migration and cancer cell invasion [41]. One consequence of this fact is that treatment by MMP inhibitors was found to be of low effectiveness against cancer metastases.

In order to gain a deeper understanding of amoeboid motion in complex environments, we focused in this work on cellular navigation between obstacles. We first examined the influence of noise on cellular motion. In our stochastic compass model, the internal direction (compass) of the cell reflects the external gradient, with some added noise. The noise level scales inversely with the external gradient strength. This model feature is based on experimental results [42,43], showing increasing CI with increasing gradient steepness. Different noise levels can also result from internal cellular characteristics. For example, normal cells exhibit a stronger response (i.e. less noise, higher motility and higher CI) to specific growth factors compared to cancerous cells of the same cell line [24]. This difference in gradient sensing between normal and cancer cells can therefore influence their ability to migrate, navigate and invade. By examining obstacle circumvention of amoeboid cells we show that cells can easily bypass obstacles of roughly their own size. This is a result of the noisy extension-retraction dynamics of membrane protrusions, which is the main characteristic of amoeboid motion.

To challenge the cell's navigation ability, we placed the cells in a maze with contradictory cues. In most biological systems, the signaling molecules can diffuse through small pores in the tissue, while the larger cells need to bypass the obstacles, for example those posed by the ECM, as described above. This is the rationale behind our choice of a signal that can diffuse through the maze walls, rather than a signal that can only diffuse via the maze openings. This is an important point, as the independent signal and maze structure pose the challenges of dead ends and corners, which the cell needs to overcome. Conversely, when the signal diffuses only in the maze, the cell merely needs to follow the chemoattractant concentration [44]. In some recent work of Sasai, a somewhat similar challenge was posed, using a dead end that forced the cell to change its direction. However, no long-term success rates were measured, and the effects of memory were not investigated.

Our maze simulations show that adaptive noise is insufficient for efficient and successful navigation. In the different maze/signal configurations that we studied, success rates varied from 0% (two cases) to 29.9%, which means that most of the cells were unable to successfully exit (or solve) the maze. These low success rates suggest the hypothesis that real cells have more "intelligent" ways to find their way than by simply obeying the external signal. After adding a simple memory effect by a secreted chemical, success rates in three separate mazes rose to 72%, 76% and 99%, respectively. Our results should be thought of as "proof of concept" that a simple self-employed memory effect can indeed improve navigation ability significantly. Since our maze was arbitrarily created, any similar maze with a contradictory chemotactic signal should lead to qualitatively similar results.

Secreted and diffusible chemicals can be found in many biological systems, for example in quorum sensing bacteria. The bacteria secrete pheromones that diffuse in the colony and are detected by the cells themselves, initiating, for example, stress response in a crowded colony. In our simulation the chemical similarly diffuses out of the secreting cell and in the surrounding environment. While no specific biological components can yet be identified with this hypothesized memory mechanism, some cells are known to leave chemical traces behind them as they move [16,17]. In [45], a migrating neutrophil encountered a path bifurcation, with two different chemoattractant levels. While the first neutrophil chose the path of higher concentration, the following neutrophil avoided that path and surprisingly chose the other one. This behavior suggests that the presence of a neutrophil masks the chemoattractant, thus effectively redirects the second cell. In the case of metastasizing cancer cells, a chemorepellent secreted by the cells may explain the broad spatial distribution of cells that is typically seen in the tissue. At the higher level of multi-cellular organisms, such as the carabid beetle, conspecific avoidance mechanisms have been identified [46]. This mechanism is believed to lead to better exploration of sparsely and randomly distributed prey resources

The existence of cells coping with both chemical signaling and environmental barriers was demonstrated in the embryo of medaka fish. Macrophages were often constrained within relatively flat, peripheral zones [47], possibly due to high tissue density or adhesiveness. However, macrophages were able to respond to a wound signal while still respecting the tissue barriers, by taking a longer path through areas that were easier to invade. How the two contradicting signals (moving towards the wound and dealing with barriers) are balanced in this example is currently unknown.

Our predictions could easily be tested by creating maze-like geometries and allowing cells to migrate therein. In fact, two very recent reports have shown how this can be done. In [5], a simple set of path bifurcations were presented to neutrophils moving under a chemokine gradient. The cells were able to successfully choose the short path. However, this study did not investigate the case where the local chemical cues are insufficient for proper navigation, and the cells indeed followed the steeper gradient. The "frustrated" situation where this simple strategy would lead to trapping is in our opinion more generic. In [46], paths were etched in a collagen matrix as a way of creating a more faithful *in vitro* analog of extracellular matrix (see also [48]); the authors then studied the migratory capabilities of cancer cells in their construct. Again, the questions of primary concern here were not specifically addressed, as in this case there was no controlled gradient providing directional information.

To summarize, we studied amoeboid motion using a computational model for cellular navigation. Our model shows that cells moving in this manner can avoid being trapped at small but not large obstacles. We then demonstrated that a simple marker strategy can improve navigation in complex terrains. This realization provides important clues into mechanisms that might be employed by real cells' migration in complex environments as well as suggests that location memory should be incorporated into robotic navigation designs. The results can be used in the study of mammalian cell migration and cancer metastasis.

## Methods

The model cell is represented by a list of connected nodes, with a set of forces acting of them. The total force on each node involves membrane protrusion force, cortical tension, friction and an effective force resulting from a constant volume constraint. The exact form of the different components is described in detail in Supporting Text S1, and the used parameter values are given in Supporting Table S1. The protruding force is localized at an active area on the membrane, which we term "patch" (see [29] for the biological context of these patches). The patch is localized at the front of the cell with respect to the external direction, as dictated by the chemoattractant gradient.

Once the total force acting on a node has been calculated, the velocity change and the resulting translocation are simply calculated using Newton's law.

# Figure Captions

Fig. 1. Model cells. The number of active protrusions varies between different cell types, and is a parameter of the model. (a)-(b) A cell with a single active protrusion (marked in red). Splitting of the front occurs when a new protrusion is created. (c)-(d) A cell with two active protrusions.

Fig. 2. Cell navigation in a gradient with no obstacles. The cell moves in a terrain with a constant gradient and no obstacles. The cell detects the chemoattractant concentration (color coded from blue-low to red-high) and moves accordingly. (a) With low noise, i.e. narrow noise distribution, the cell path is highly accurate. (b) With higher noise, i.e. wider noise distribution, the cell wanders around and its path is less accurate.

Fig. 3. Navigation between obstacles. In the two different terrains, the cell slides along the obstacle and bypasses it, according to the general direction of the gradient (color coded as in **Error! Reference source not found.**). (a) A terrain with a constant gradient towards a point source. (b) A terrain with a fixed slope.

Fig. 4. The maze. The model cell is initially placed in the "Start" position (arbitrarily chosen). The cell moves according to the chemoattractant concentration, as indicated in color code (blue-low to red-high). The arrows show the local gradient direction. In some sections of the maze, the chemoattractant gradient leads to a corner, a wall or a dead end, as shown for example by the asterisks.

Fig. 5. The cell's path in the maze. The cell's center of mass is plotted, from the maze start to the end. (a-b) When the cell is exposed to high noise (i.e. wide noise distribution) it can explore different paths and exhibits a curved trail. (c) With a lower noise level, the path is more accurate and the cell chooses the shortest path. (d) With low noise, the cell may get stuck behind a wall or a corner in the maze.

Fig. 6. A cell navigating in the maze. (a) A stuck cell: The local chemoattractant gradient points towards a maze wall. When the cell reaches the wall it moves around, as shown by the different contours, but is unable to bypass the local barrier. (b-c) A successful cell: (b) With a chemorepellent secreted and detected, the cell can overcome the barrier and continue to move until the goal is reached. (c) The chemorepellent trail: The color represents the difference between the chemoattractant and the chemorepellent, as actually detected by the cell, therefore the areas visited by the cell have a lower concentration than their vicinity.

Fig. 7. The effect of the secreted chemical on navigation success. Maze success rates for regular chemotactic navigation (no chemical secreted) and for bootstrapping navigation with different chemical diffusivities. Bootstrapping navigation improves success rates from 30% to 99%, as long as the chemical's diffusivity is not too high. Very high diffusivity blurs the spatial information and thus reduces navigation success rates. About 250 independent runs on the same maze were sampled for each case.

Fig. 8. Passage times. The time a successful cell travels from the starting point to the ending point of the maze. Bootstrapping navigation method reduces the passage time and thus improves the cell's performance, as long as the chemical's diffusivity is not too high. When the chemical diffuses too fast, the spatial information is blurred and the passage time increases. About 250 independent runs on the same maze were sampled for each case.

# Supporting information

Supporting Text S1. Amoeboid motion: model description and computational details.
Supporting Table S1. Parameter values.

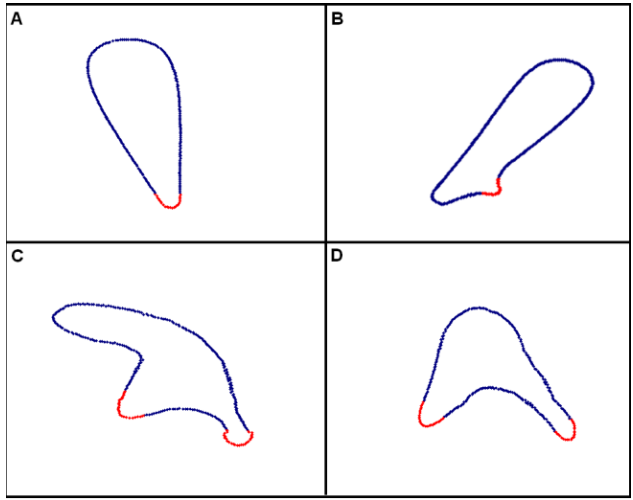

Fig. 1.

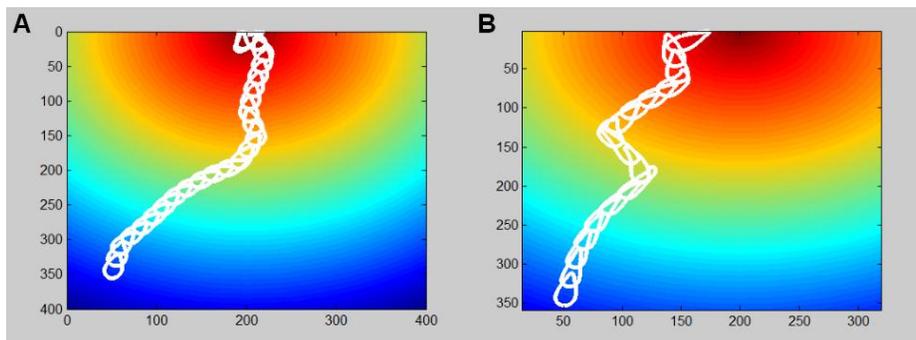

Fig. 2.

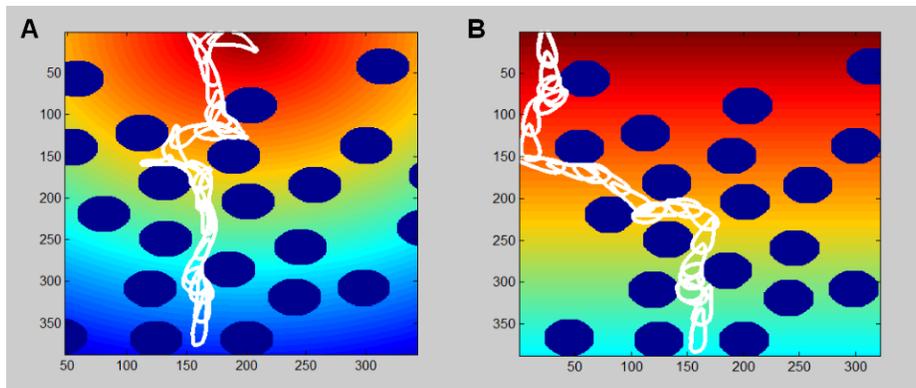

Fig. 3.

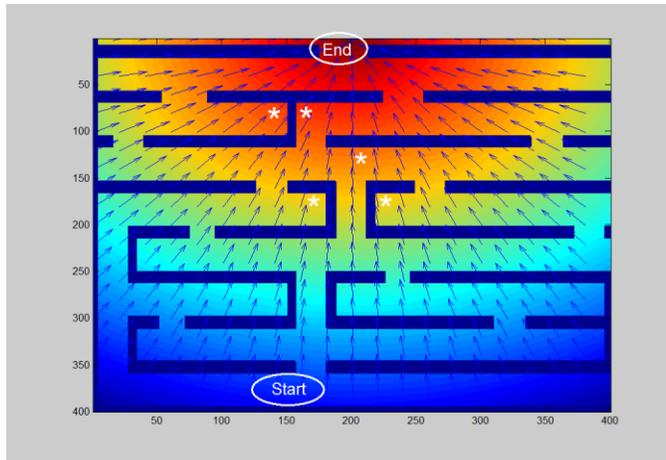

Fig. 4.

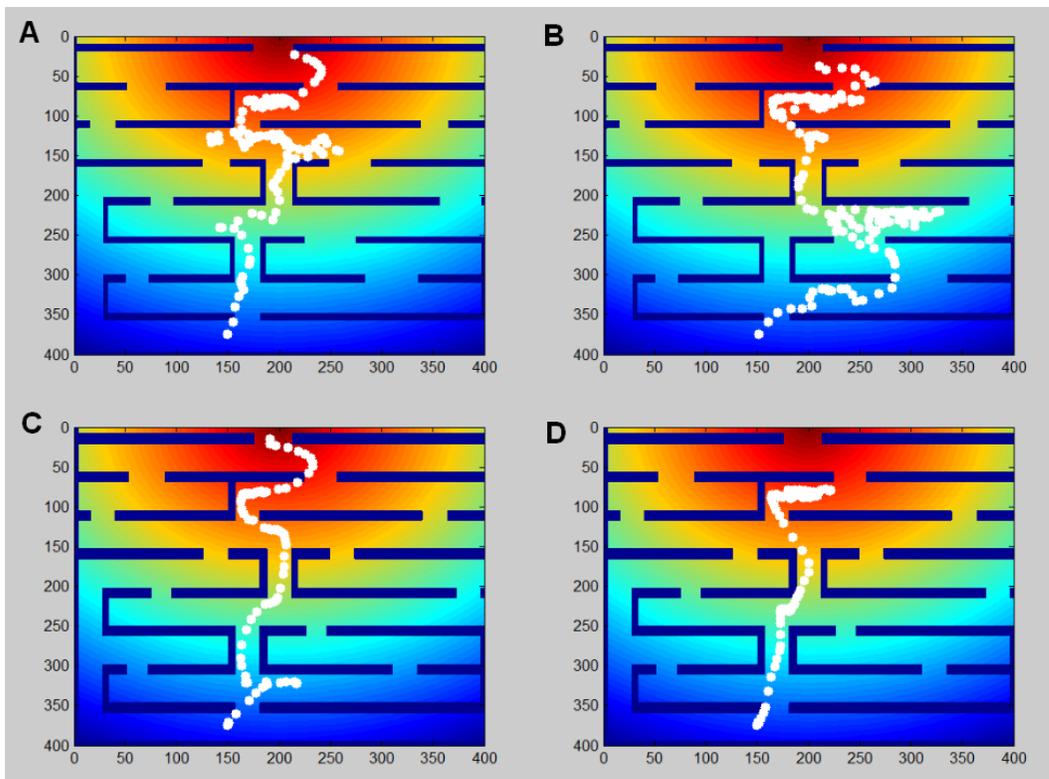

Fig. 5.

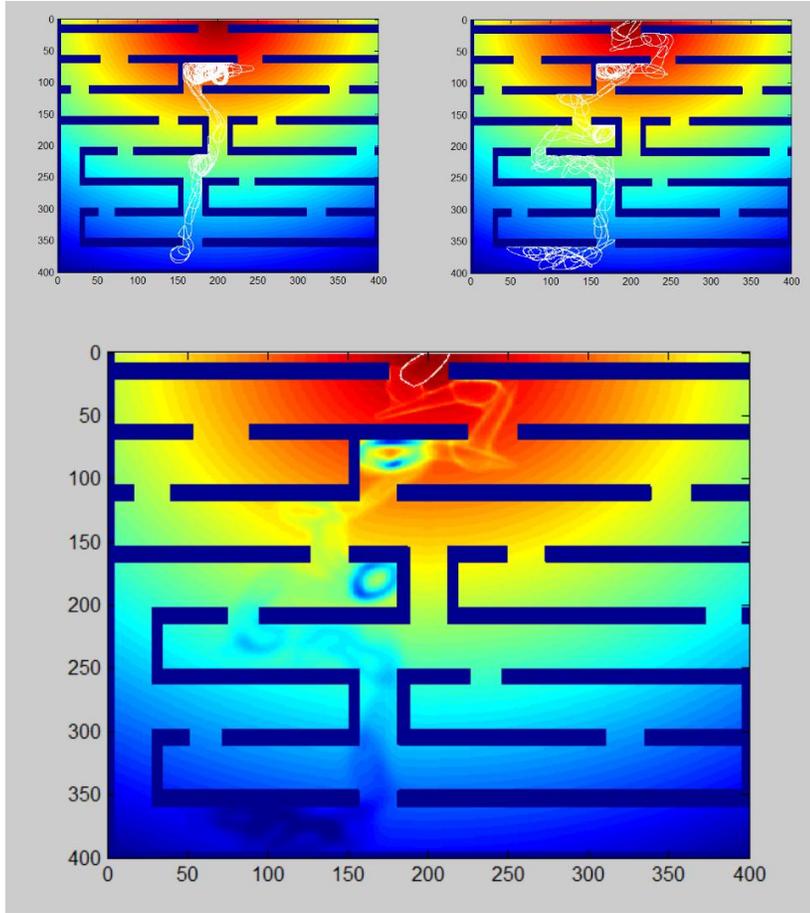

Fig. 6.

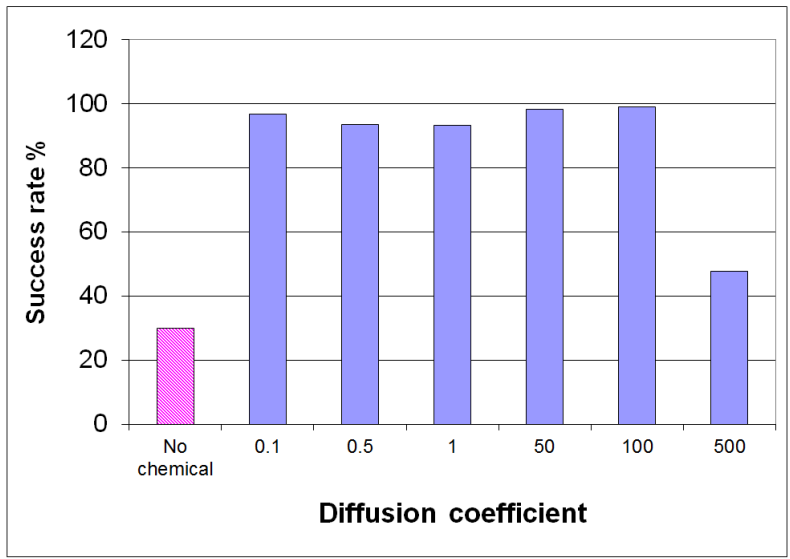

Fig. 7.

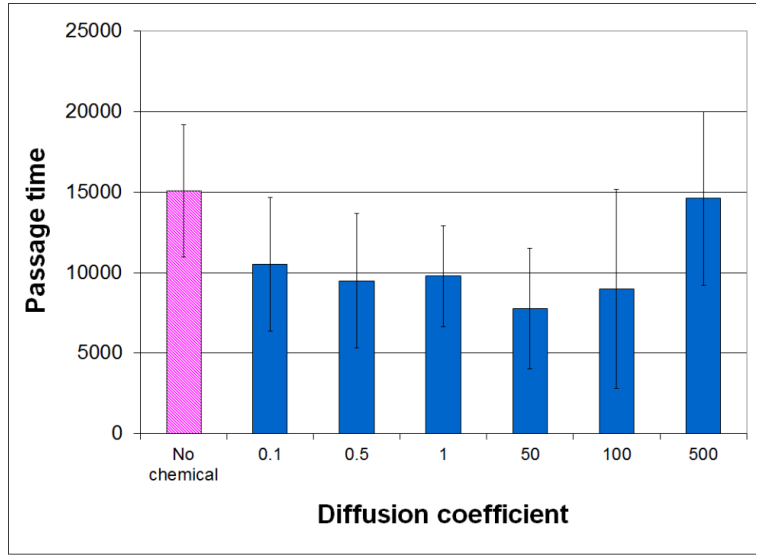

Fig. 8